\documentstyle[aps,floats,graphicx]{revtex} 
\twocolumn

\newcommand{\cH}{{\mathcal H}}
\newcommand{\cP}{{\mathcal P}}

\DeclareGraphicsExtensions{.eps} 
\graphicspath{{figures/}{./}}
\newcommand{\myfig}[1]{\begin{figure} 
			  \begin{center} 
		\includegraphics*[width=0.9\linewidth]{#1}
                        \end{center}}

\begin{document}
\wideabs{ 
\title{Spin-$1$ chain doped with mobile $S=1/2$ fermions}
\author{Beat Ammon\cite{ABA}
and Masatoshi Imada\cite{AMI}}
\address{ISSP, University of Tokyo, 7-22-1 Roppongi, Minato-ku, Tokyo 106,
	 Japan}
\maketitle
\begin{abstract}
We investigate doping of a two-orbital chain with mobile $S=1/2$
fermions as a valid model for $\rm Y_{2-x}Ca_xBaNiO_5$. The $S=1$
spins are stabilized by strong, ferromagnetic (fm) Hund's rule
couplings. We calculate correlation functions and thermodynamic
quantities by DMRG methods and find a new hierarchy of energy scales
in the spin sector upon doping. Gapless spin-excitations are generated
at a lower energy scale by interactions among itinerant polarons
created by each hole and coexist with the larger scale of the gapful 
spin-liquid background of the $S=1$ chain accompanied by a finite
string-order parameter.
\end{abstract}
 }   


Since Haldane's famous conjecture about the existence of a spin-gap in
one-dimensinal integer Heisenberg (HB) spin-chains \cite{Haldane}, the
intriguing physics of $S=1$ chains and $S=1/2$ ladders has been
investigated in numerous theoretical and experimental studies and
especially hole doping of a spin-liquid is of great importance for the
Mott-transition in the context of high-$T_C$
superconductivity. However, doping of a Haldane $S=1$ chain with
mobile holes has only recently been realized in the material $\rm
Y_2BaNiO_5$ \cite{YBaNiO}. In the undoped case, the two active
$\rm Ni^{2+}$ orbitals are coupled by a strong fm Hund's rule coupling
$J_H$, giving an almost ideal $S=1$ HB chain with a large spin gap
of $\Delta_{S=1}\approx 100K$ and very weak inter-chain
interactions. By substituting $\rm Ca^{2+}$ by $\rm Y^{3+}$, mobile
holes can be introduced in the chain. The most remarkable experimental
features are a reduction of the resistivity $\rho_{dc}$ by several
orders of magnitude, and the creation of states with $S$ between $1$
and $3/2$ per impurity inside the Haldane gap \cite{DiTusa}. Further
the temperature dependence of the resistivity no longer follows
thermal activation over a barrier.

While doping of $S=1$ HB chains by static as well as bond impurities
\cite{Kaburagi,LuSu} has been studied by several authors, only little 
is known about doping of mobile $S=1/2$ fermions in $S=1$ chains. A
first theoretical investigation limited to the case of weak hopping
amplitudes \cite{Penc} found states inside the Haldane gap with $S$
larger than $S=1/2$ per impurity, whereas other studies based on the
weak coupling approach find that the spin gap is immediately destroyed
upon doping for models with a level difference and fully mobile
electrons \cite{Fujimoto}. Numerically, the spin dynamics of mobile
holes in $\rm Y_2BaNiO_5$ have been investigated by exact
diagonalization of small finite chains for effective one
\cite{Dagotto} and two band models \cite{Batista}, and based on the
assumption of infinitely strong Hund's rule couplings $J_{H}$ states
below the Haldane gap have been found.  

Among the most interesting properties of these systems is a
competition between fm ordering induced by the double exchange (DE)
mechanism and antiferromagnetic (af) order due to direct
exchange. Other problems are the persistence of the spin-gap and the
characterization of the doped phase. In this Letter we would like to
address for the first time this competition in an unperturbative
manner, with finite values of the Hund's rule coupling and fully
mobile electrons for system sizes significantly exceeding the
correlation length of the undoped system. We do so by investigating
groundstate properties and various correlation functions by the
density matrix renormalization group (DMRG) method \cite{White}, and
thermodynamic quantities by the thermodynamic DMRG (TDMRG)
\cite{TDMRG}. This reveals a new hierarchy of energy scales in the
spin sector with gapless spin-modes given by the polaron-polaron
interactions and a second, much higher lying energy scale stemming
from the gapful spin-liquid background of the $S=1$ chain. We also
give first numerical evidence that the spin-gap is immediately
destroyed upon doping of mobile electrons in a $S=1$ chain with level
difference. The hierarchy structure found for doping of static holes
into the spin-gap background \cite{IinoImada} surprisingly persists
even in the metallic phase with fully mobile hole doping.

We investigate a model with strong Hund's rule coupling $J_H<0$ and a
fully occupied lower band, as the $d_{x^2-y^2}$ orbibtal of $\rm
Ni^{2+}$ in $\rm Y_{2-x}Ca_xBaNiO_5$ is almost localized. The results
for a model with equal doping in both bands will be published
later\cite{Symmetric}. Further we assume af couplings $J>0$ between
neighboring sites in the upper ($3d_{z^2-r^2}$) orbital and $J_d>0$
between neighboring sites in different orbitals. Thus our Hamiltonian
reads
\begin{eqnarray} \label{ASHamil}
\cH_{\text{as}} & = & 
   -t \sum_{j,\sigma} \cP \left(
        c^{\dagger}_{j,1,\sigma} c_{j+1,1,\sigma}
        +H.c.\right) \cP \nonumber \\ 
&& 
+ J \sum_{j} \left(\vec{S}_{j,1} \vec{S}_{j+1,1}
		-{1\over4}n_{j,1}n_{j+1,1}\right)
\nonumber \\ 
  &&
   + J_d \sum_{j} \left(
		     \vec{S}_{j,1} \vec{S}_{j+1,2} + 
		     \vec{S}_{j,2} \vec{S}_{j+1,1}
   		     -{1\over2}n_{j,1} 	 \right)   
\nonumber \\ 
&&  
+J_H \sum_{j} \vec{S}_{j,1} \vec{S}_{j,2}
  - \mu_1 \sum_{j} n_{j,1} \nonumber,
\end{eqnarray}
where the first index of the double index $j,i$ denotes the lattice
site and the second index the orbital, the projection operator $\cP$
prohibits double occupancy of a site in order to take account of the
strong on-site Coulomb repulsion and the rest of the notation is
standard. If not otherwise mentioned we always set $J=J_d$ and assume
$J_H\gg J,J_d,t$, typically $-J_H=10t=20J=20J_d$. We have also
performed calculations with two different chemical potentials for the
upper and lower band, but we find finite hole densities in the lower
band only at the highest temperatures $T>t$ in that case, and the
results are identical to the above case for $T<t$.

The TDMRG allows us to calculate thermodynamical properties in the
thermodynamic limit of infinite system size. In general we use finite
Trotter-time step sizes of $\Delta \tau=0.2t$ and extrapolate from the
grand canonical ensemble with fixed chemical potential to constant
particle density first. We keep between $60$ and $80$ states per
system and environment block, and use a re-biorthogonalization method
\cite{BiOTDMRG} if we encounter numerical instabilities. With the
groundstate DMRG we have considered system sizes of up to $2\times
256$ sites and kept up to $1300$ states per system and environment
block.

We begin by considering two limits which are easy to understand. The
first case is $J=J_d=0$. Away from half filling, all spins are
ferromagnetically aligned by the DE mechanism \cite{DblExch} in order
to gain kinetic energy. The second case is at half filling, where upon
switching on the af couplings $J,J_d$, the system can be mapped to an
af $S=1$ HB chain, with a spin-gap and exponentially decaying af
spin-correlations. For any finite value of $J,J_d$, and finite hole
doping we will therefore have competition between fm order induced by
the DE mechanism and the spin liquid state of the $S=1$
chain. Let us briefly sketch the picture that emerges from our
numerical calculations, before investigating the physical quantities
in detail. Each hole doped into the $S=1$ chain is surrounded by a
small fm cloud created by the DE mechanism, and we will
call this a polaron in the following. However, in the parameter
regions we studied in this paper, the polaron is only a weak local
perturbation of the spin-liquid background, and the correlation length
remains close to its original value. Among the polarons, $4k_F$ and
$2k_F$ charge density wave (CDW) order is then stabilized, and the
lowest lying spin excitations are weak magnetic interactions among the
polarons. Due to these interactions, there is a spin-singlet
groundstate already for very weak couplings $J,J_d
\gtrsim 0.2t$ and $J_H=-10t$.

Let us turn to the detailed investigation of physical quantities next
and address the question of whether the spin-gap is robust upon doping
of holes first. The mapping to the af $S=1$ HB chain at
half filling gives effective couplings $J_e=3/4 J$ of the original
$S=1$ chain by perturbation theory, hence we expect a spin gap of
$\Delta_s=3/4\Delta_{S=1}$ of the original value $\Delta_{S=1}\approx
0.41J$. In good agreement we obtain $\Delta_s \approx 0.153 t\pm 0.005
t = 0.306 J \pm 0.01J$ from the TDMRG by $\lim_{T\rightarrow 0}
\sqrt{2Tc_v/(3\chi)} \rightarrow \Delta_s$ \cite{Xiang}. Alternatively, 
we have calculated the spin gap by finite size scaling
$\Delta_s=\lim_{T\rightarrow 0} \Delta_{s}(L;N=Ln)$, where
$\Delta_{s}(L;N=Ln)=\Delta_{s}(L;N)=E_0(L;N;S^z=1)-E_0(L;N;S^z=0)$ and
$E_0(L;N;S^z)$ is the groundstate energy of the system with $N$
particles on $L$ sites and total $S^z$ component of the spin along the
$z$-direction. However care needs to be taken in order to avoid free
$S=1/2$ spins at the end of the open chain. To keep the free spins
out, we have imposed the boundary condition that the upper orbital on
the boundary sites remains empty. The result for different values of
the hole concentration in the conduction band $n_h^c$ is listed in
Tab.~\ref{Tab0TSpin} and agrees well with the TDMRG.
\begin{table}[b]
\begin{tabular}{l|c|c|c}
$n_{h}^c$ & $\Delta_s$ 
		& $v_{\sigma}$ &$\chi^{0}$ \\
\hline
0 & $0.1504 t$ & - & 0 \\
0.0625 & 0 & 0.244 & 2.61 \\
0.125 & 0 & 0.201 & 3.17 \\
\end{tabular}
\caption[*]{The spin-gap $\Delta_s$ for different values of hole 
	doping $n_h^c$ in the conduction band and
	$-J_H=10t=20J=20J_d$. Also listed is the spin-velocity
	$v_{\sigma}$ and the susceptibility $\chi^0$ at $T=0$. The results
	are obtained after extrapolations to the thermodynamic limit
	by finite-size scaling.}
\label{Tab0TSpin}
\end{table}
\myfig{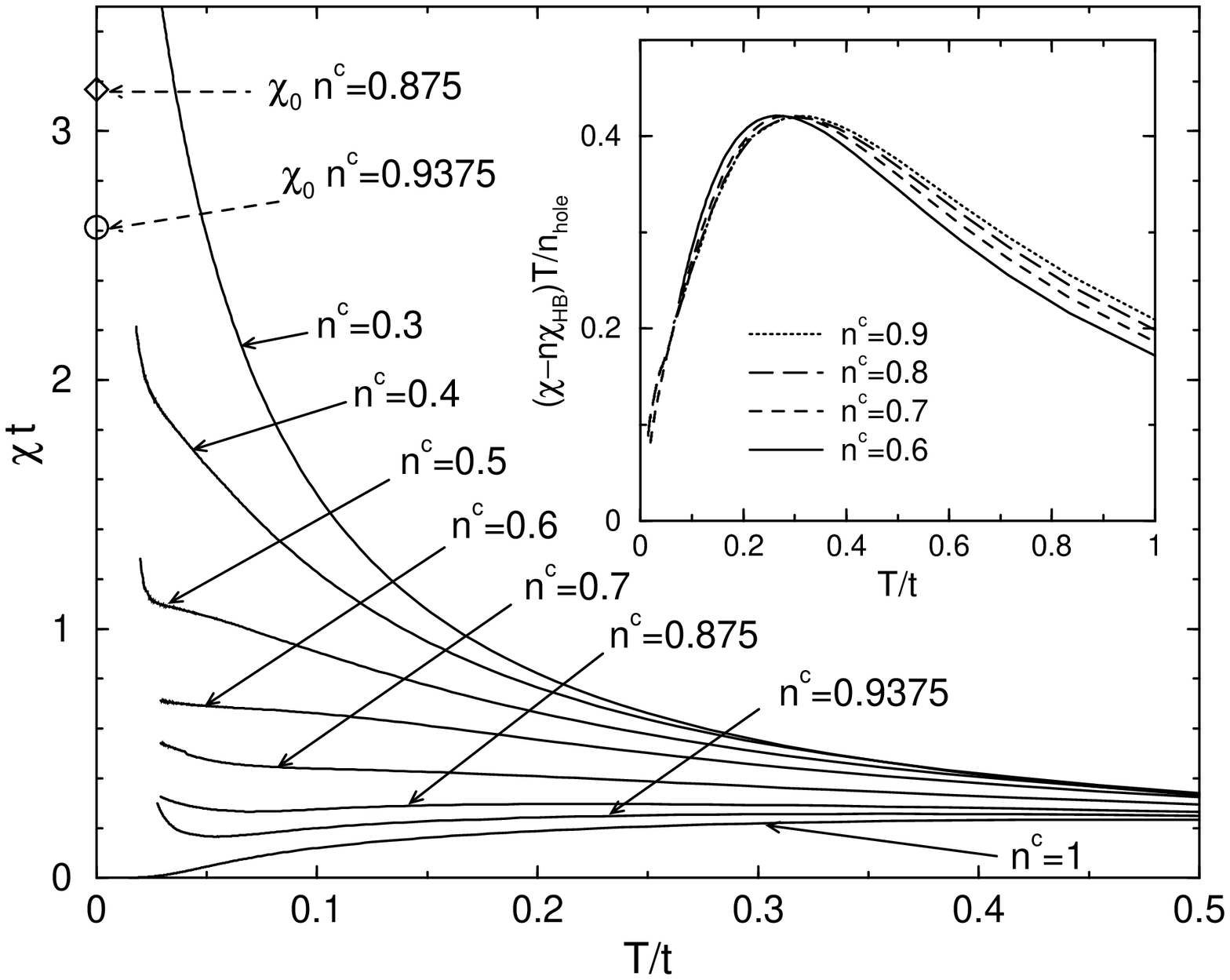}
   \caption[*]{Magnetic susceptibility $\chi$ for $-J_H=10t=20J=20J_d$.
	Inset: Effective Curie-constant $S(S+1)/3$ per hole.}
\label{FigChi}
\end{figure}
Here we see that the spin-gap is immediately destroyed upon doping,
and by noting that the smallest momentum in the finite open chain is
$k_{\text{min}}=\pi/L$ we can calculate the spin-velocity $v_\sigma$
from $\Delta_s(L)=v_{\sigma}k_{\text{min}}$. Thus we can determine the
$T=0$ susceptibility by $\chi^0=\frac{2K_\sigma}{\pi v_\sigma}$, where
$K_\sigma=1$ due to the SU(2) symmetry. Both results are listed in
Tab.~\ref{Tab0TSpin}. We compare these findings with the TDMRG results
for $\chi(T)$ in Fig.~\ref{FigChi}. The strong enhancement of
$\chi(T)$ at low temperatures suggests the formation of larger
magnetic moments upon doping, in agreement with the $T=0$ results. In
the inset of Fig.~\ref{FigChi} we estimate the effective
Curie-constant $\gamma\equiv S(S+1)/3$ created by each hole by
subtracting the background of the $S=1$ chain $\gamma=(\chi-(1-n^c_h)
\chi_{S=1}) T / n^c_h$. The maximum value of $\gamma\approx 0.5$
near $T=0.3t$ is nearly independent of the hole doping and corresponds
to $S_{\text{eff}}\approx 0.7$, in rough agreement with experimental
findings \cite{DiTusa,Kojima}. By further lowering the temperature, af
interactions among the polarons reduce $\gamma$ and finally lead to
a singlet groundstate at $T=0$. We have checked the rotational
invariance by the DMRG from $\sigma(i,j)=S_i^z S_j^z-1/2(S_i^+ S_j^-)$
which vanishes for a singlet groundstate and confirm this result within the
numerical precision of the DMRG ($|\sigma(i,j)|<10^{-6} \forall i,j$).

%
Next we investigate correlation functions by the DMRG method and start
with the particle density in the conduction band $n^c(x)$, which shows
Friedel oscillations induced by the open boundary conditions. For small hole
densities $n_h^c=1/16$, and $n_h^c=1/8$ we only find $L n_h^c/2$ hole
pockets (see inset Fig.~\ref{ChDensFig}) indicating hole pairing. For
larger hole densities however, a weak structure in the hole
pockets suggests that the two holes bound to a pair are separated by
several lattice spacings. The pair binding energy in the low hole
density region can be estimated by $\Delta_{\text{pair}}=2 E_1-E_0-E_2
\approx 0.016t$ for $-J_H=10t=20J=20J_d$, where $E_n$ is the
groundstate energy with $n$ holes, and confirms pair formation. Also
note the very weak amplitudes of the Friedel oscillations.
\myfig{fig2}
   \caption[*]{Fourier transforms $n(k)$ of the charge density 
	$n^c(x)$ for hole concentrations of $n_h^c=0.0625$ and $n_h^c=0.125$ 
	in the conduction band and $-J_H=10t=20J=20J_d$. Inset: charge
	density $n^c(x)$ for $n_h^c=0.125$.}
\label{ChDensFig}
\end{figure}
In the Fourier-transform $n(q)$ we only find one peak for very low
hole doping $n_h^c=1/16$ at $k=n_h^c \pi$ and two peaks at $k=n_h^c
\pi$ and $k=2 n_h^c \pi$ for larger hole doping $n_h^c=1/8$.
In general, a Tomonaga-Luttinger liquid shows $2k_F$ and $4k_F$
fluctuations, and $2k_F=(n^c+1)\pi$ for a large Luttinger volume
involving the electrons in the lower band. Our observation is
consistent with these $2k_F$ and $4k_F$ periods for a large Luttinger
volume. By fitting to the Friedel-oscillations for an impurity potential
$\delta_n(x)\propto C_1 \cos(2k_F x) x^{-(1+K_\rho)/2} + C_2 \cos(4k_F
x) x^{-2K_\rho}$ \cite{Fabrizio} we have determined the single
correlation exponent $K_\rho\approx 0.51\pm 0.05$ for $n_h^c=1/8$,
indicating dominant CDW correlations. Because of trapped states near
the boundaries, these sites have to be discarded for the fit
and the uncertainty stems from the fitting ambiguity.

Independently, the correlation exponent $K_{\rho}$ can also be determined
from the pairing correlations $P_{i}(f)P_{i+r}^\dagger(f)$, where 
$P_{i}^\dagger(f)=\frac{1}{\sqrt{2}}(c_{i,\uparrow}^\dagger 
	c_{i+f,\downarrow}^\dagger \mp c_{i,\downarrow}^\dagger 
	c_{i+f,\uparrow}^\dagger)$ is either the singlet ($-$)
or triplet pair ($+$) creation operator. The singlet pairing
correlations decay as $P_{i}(f)P_{i+x}^\dagger(f) \propto
A_0\ln(x)^{-1.5}x^{-1-1/K_\rho} +A_2\cos(2k_Fx)x^{-K_\rho-1/K_\rho}$,
and in agreement with the previous estimate we obtain $K_{\rho}\approx
0.51 \pm 0.05$ by simultaneously fitting to $f=2,4,6$, and $8$ (see
Fig.~\ref{PairingFig}). Note that the largest amplitudes of the
pairing-correlation are obtained for $f=6$ and $8$, in agreement with
the pair structure in $n(x)$.
\myfig{fig3}
   \caption[*]{Singlet pairing correlations
   $P_{i}(f)P_{i+x}^{\dagger}(f)$ with $P_{i}^\dagger(f)=
   \frac{1}{\sqrt{2}}(c_{i,\uparrow}^\dagger c_{i+f,\downarrow}^\dagger
   -c_{i,\downarrow}^\dagger c_{i+f,\uparrow}^\dagger)$ 
	for $-J_H=10t=$ $20J=20J_d$ and $n^c_h=1/8$.}
\label{PairingFig}
\end{figure}
\myfig{fig4}
   \caption[*]{Spin-spin correlations $S^z_iS^z_j$ measured
	from the center and end of the chain for different values of 
	hole doping with $-J_H=10t=20J=20J_d$. Inset: string-correlation
	function g(x).}
\label{SpinSpinFig}
\end{figure}

Finally we turn to the spin-spin correlations shown in
Fig.~\ref{SpinSpinFig}. As for the undoped case, all spin-spin
correlation functions seem to decay exponentially
$S^z_iS^{z}_{i+x}\propto \cos(2k_F x) e^{-x/\xi}$, and the correlation
length $\xi=7.9\pm 0.1$ for $n_h^c=1/16$ and $\xi=11.2\pm 0.2$ for
$n_h^c=1/8$ remains close to the undoped case $\xi\approx 6.1$.
\begin{figure}[t]
\begin{center} 
	\includegraphics*[width=.9\linewidth]{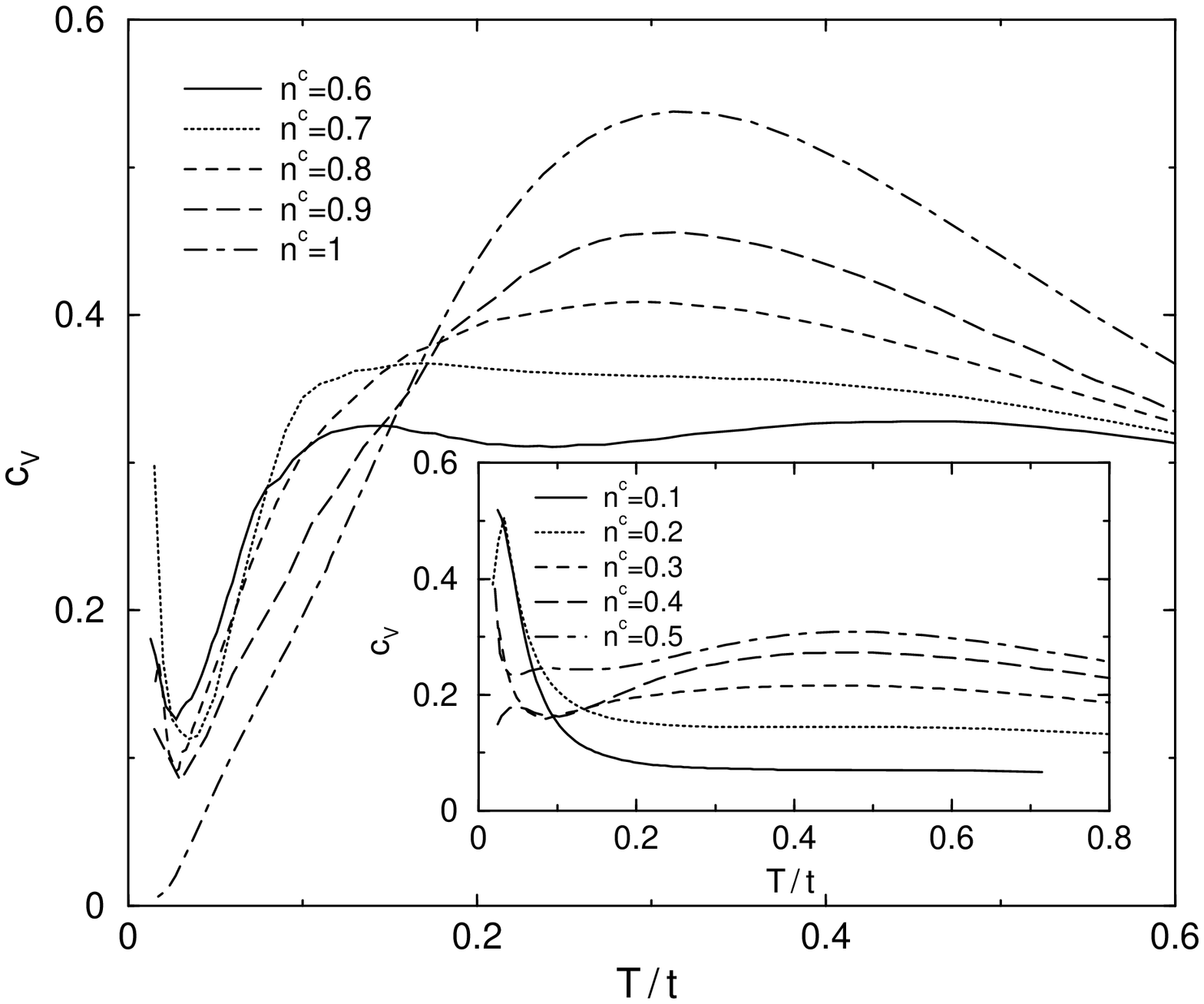}
   \caption[*]{Specific heat $c_V$ calculated by the TDMRG
	 for $-J_H=10t=20J=20J_d$.}
\label{SpecHeatFig}
\end{center}
\end{figure}
A more conclusive quantity is the string order parameter
\cite{StringOrder} $g(x)=\langle (\sum_{i=1,2} S_{x_0,i}^z) \left(
\prod_{k=x_0+1, j=1,2}^{x-1} e^{i \pi S_{k,j}^z} \right)
(\sum_{l=1,2} S_{x,l}^z)\rangle $ characterizing the Haldane $S=1$
chain. The finite value of $g(x)$ upon doping suggests that the
spin-liquid state remains intact. However, for the doped case it is not
clear whether $g(x)$ remains finite in the thermodynamic limit,
because our system sizes do not seem to be long enough if
polaron-polaron correlations would destroy the string correlation.

We finish by summarizing how all the findings including the surprising
coexistence of gapless spin-modes with exponential spin-correlations
(at intermediate distances) can be reconciled within the polaron
picture. The holes which create a small fm cloud by the DE mechanism
are bound to pairs as we see from the charge density, and $2k_F$ and
$4k_F$ CDW correlations are dominant with $K_\rho\approx 0.51$. The
amplitudes of the CDW are very weak (see Fig.~\ref{ChDensFig}). This
and the finite string-order suggest that the polarons are only a weak
perturbation of the spin-liquid background of the $S=1$
chain. Therefore we propose a new model with two very different
energy-scales for the spin-excitations. The larger one is of the order
of $\Delta_s$ and stems from the spin-liquid background, giving rise
to exponentially decaying spin correlations at the distances we can
observe by the DMRG. The second energy scale is very small and
consists of the af interactions among the polaron-pairs. These are the
gapless, lowest lying spin excitations, and at very long distances we
expect power-law decay of the spin-correlation functions. Since these
interactions are among polaron-pairs, even our largest system-sizes
contain only few such pairs and are too small to discriminate between
exponential and power-law decay. Note however that $\xi_{\text{spin}}$
increases with doping. In order to estimate the energy scales we
finally show the specific heat $c_V$ calculated by the TDMRG in
Fig.~\ref{SpecHeatFig}. From the position of the large broad peak near
$T\sim 0.3t$ we see that the energy-scale of the spin-liquid remains
almost unchanged upon doping, and from the sharp increase of $c_V$ at
the lowest temperatures we estimate magnetic interactions among
polarons at an energy scale of $T \sim 0.02J$.

We wish to thank H. Asakawa, N. Shibata, M. Sigrist, and H. Tsunetsugu
for valuable discussions. The numerical calculations have been
performed on workstations at the ISSP.


%
\end{document}